\documentclass[reprint,amsmath,amssymb,aps,citeautoscript,footinbib]{revtex4-1}

\pdfoutput=1

\usepackage{floatrow}
\usepackage{graphicx}
\usepackage{epstopdf,subfig,epsfig}
\usepackage[export]{adjustbox}
\usepackage{dcolumn}
\usepackage{bm}
\usepackage{marginnote}
\usepackage{changebar}
\usepackage{caption}
\usepackage{natbib}
\usepackage[colorlinks]{hyperref}

\hypersetup{
    colorlinks=true,       
    linkcolor=blue,        
    citecolor=blue,        
    filecolor=black,       
    urlcolor=black,        
    bookmarks=false,
    pdffitwindow=true,
    pdfpagelayout=SinglePage,
    plainpages=false
}

\graphicspath{{figures/}
              {../numerics/GP/}
             }

\renewcommand{\i}{\mathrm{i}}

\renewcommand{\d}{\mathrm{d}}

\DeclareMathOperator{\sgn}{sgn}

\newcommand{\const}{\mathrm{const}}

\begin{document}

\title{Nonlinear Schr\"{o}dinger equation in cylindrical coordinates}

\author{R. Krechetnikov  (\url{krechet@ualberta.ca})}
\affiliation{University of Alberta, Edmonton, Alberta, Canada T6G 2E1}

\date{\today}

\begin{abstract}
Nonlinear Schr\"{o}dinger equation was originally derived in nonlinear optics as a model for beam propagation, which naturally requires its application in cylindrical coordinates. However, the derivation was done in the Cartesian coordinates with the Laplacian $\Delta_{\perp} = \partial_{x}^{2} + \partial_{y}^{2}$ transverse to the beam $z$-direction tacitly assumed to be covariant. As we show, first, with a simple example and, next, with a systematic derivation in cylindrical coordinates, $\Delta_{\perp} = \partial_{r}^{2} + \frac{1}{r} \partial_{r}$ must be amended with a potential $V(r)=-\frac{1}{r^{2}}$, which leads to a Gross-Pitaevskii equation instead. Hence, the beam dynamics and collapse must be revisited.
\end{abstract}

\maketitle

The 2D nonlinear Schr\"{o}dinger equation (NLS) for the envelope $\psi$ of a quasi-monochromatic light wave
\begin{align}
\label{NLS:basic:non-dimensional}
\i \psi_{\tau} + \Delta_{\perp} \psi + \beta |u|^{2} \psi = 0,
\end{align}
where $\Delta_{\perp} = \partial_{x}^{2} + \partial_{y}^{2}$ is the Laplacian transverse to the beam $z$-direction, was originally derived \cite{Kelley:1965,Talanov:1965} for nonlinear Kerr media. With an equivalent formulation, when time $\tau$ is replaced by $z$, it was used in a number of works \cite{Kelley:1965,Talanov:1965,Zakharov:1976b,Zakharov:1971,*Sobolev:1973,*Afanasjev:1995,*Soljacic:1998,*Akhmediev:1998} to name a few, while the form \eqref{NLS:basic:non-dimensional} was employed in most other studies, e.g. \cite{Lomdahl:1980,Papanicolaou:1982,Wood:1984,*Sulem:1984,Rypdal:1986,Sulem:1999,Fibich:2015,*Ablowitz:2019}. Besides fiber optics, in which NLS is a central model for high bit-rate communication \cite{Ablowitz:2011}, it also found a myriad of other applications such as in laser-tissue interaction \cite{Powell:1993,*Fibich:1995,*Fibich:1996}, plasma physics \cite{Zakharov:1972b}, wave focusing regions of the ionosphere \cite{Gurevich:1978}, and energy transport along molecular chains \cite{Davydov:1981}. Equation \eqref{NLS:basic:non-dimensional} attracted a lot of attention due to its rich properties, in particular, because it proved to be a paradigmatic example of inverse scattering transform \cite{Zakharov:1972}. Since NLS \eqref{NLS:basic:non-dimensional} is known to lead to a finite-time singularity \cite{Glassey:1977}, of key interest has been the extensive research on the rate at which the singularity is approached starting with \cite{Kelley:1965,Zakharov:1976b}; see overview of the vast literature in \cite{Rypdal:1986,Sulem:1999}. A closely related question is on beam self-focusing exhibited in \eqref{NLS:basic:non-dimensional}, naturally studied in a cylindrical system of coordinates with radial symmetry by taking $\Delta_{\perp}$ to be the radial Laplacian $\Delta_{r} = \partial_{r}^{2} + \frac{1}{r} \partial_{r}$ \cite{Chiao:1964,Zakharov:1971,Zakharov:1976b, Papanicolaou:1982,Wood:1984,Rypdal:1986,*Jones:1988}; in this case we will call \eqref{NLS:basic:non-dimensional} \textit{a radial NLS}. Since \eqref{NLS:basic:non-dimensional} was originally derived in the Cartesian coordinates \cite{Kelley:1965,Talanov:1965,Newell:1992}, the transition to the cylindrical system is then implicitly performed based on the principle of covariance, i.e. independence of physical laws of a coordinate system. However, despite the ``universal'' character, NLS does not represent fundamental physical laws, but rather a long-wave asymptotic reduction of the latter such as Maxwell's equations.

To understand the irrelevance of the covariance principle, let us start with a heuristic derivation \cite{Debnath:2012,*Vitanov:2013} of the linear part of the envelope equation \eqref{NLS:basic:non-dimensional} in cylindrical coordinates, e.g. for the free surface elevation $\eta(t,r)$ of axisymmetric 2D water waves, which can be naturally represented in terms of Hankel harmonics:
\begin{align}
\eta(t,r) = \int_{0}^{\infty}{\widehat{\eta}_{0}(k) J_{0}(k r) e^{-\i \omega(k) t} \, k \d k} + \mathrm{c.c.},
\end{align}
where $\widehat{\eta}_{0}(k)$ is the Hankel transform of the initial free surface elevation. The asymptotic expansion of this expression away from the origin, $k r \gg 1$, and in the form of a narrow wavepacket $\delta k = \epsilon \kappa$ of width $\epsilon$ near some fixed wavenumber $k$ yields $\eta(t,r) = e^{\i (k r - \omega t)} (\epsilon/R)^{1/2} \, \psi(T,\tau,R) + \mathrm{c.c.}$, that is a traveling wave $e^{\i (k r - \omega t)}$ modulated with an envelope function
\begin{align}
\label{eqn:envelope}
\psi \sim \int_{-\infty}^{\infty}{\widehat{\eta}_{0}(k,\kappa) e^{\i \left(\kappa R - \dot{\omega}(k) T - \frac{\ddot{\omega}(k)}{2} \kappa^{2} \tau - \frac{\pi}{4}\right)} \, \kappa^{1/2} \d \kappa}
\end{align}
evolving on slow time $T = \epsilon t$, $\tau = \epsilon^{2} t$ and spatial $R = \epsilon r$ scales; here $\dot{( \ )} = ( \ )_{k}$. Differentiation of \eqref{eqn:envelope} leads to
\begin{align}
\label{eqn:SE:cylindrical}
\i \psi_{\tau} + \frac{\ddot{\omega}(k)}{2} \left(\Delta_{R} - \frac{1}{4 R^{2}}\right) \psi = 0,
\end{align}
i.e. compared to \eqref{NLS:basic:non-dimensional} the radial Laplacian $\Delta_{R}$ is appended with a potential $- \frac{1}{4 R^{2}}$. Notably, in this example the transformation $\psi = R^{-1/2} \, \widetilde{\psi}(\tau,R)$ reduces \eqref{eqn:SE:cylindrical} to the Schrodinger equation $\i \, \widetilde{\psi}_{\tau} + \frac{1}{2} \ddot{\omega}(k) \, \widetilde{\psi}_{RR} = 0$ for a ``free particle'', i.e. the effect of the potential is to modify the amplitude of the wave as it travels to/from the origin. As follows from the behavior of the Bessel function $J_{0}(k r)$, the speed of wave propagation varies as one gets closer to the origin thereby leading to a more substantial change in the wavepacket width compared to the limit $k r \gg 1$. To account for that, the potential must therefore be modified from $- \frac{1}{4 R^{2}}$, as we will see from the derivation below.

The systematic derivation of the envelope equation here is analogous to \cite{Newell:1992}, but performed in cylindrical coordinates. A starting point is the Helmholtz equation:
\begin{align}
\label{eqn:EM-propagation}
\nabla^{2} \boldsymbol{E} - \nabla \left(\nabla \cdot \boldsymbol{E}\right) - \frac{1}{c^{2}} \partial_{t}^{2} \boldsymbol{E} = \frac{1}{\varepsilon_{0} c^{2}} \partial_{t}^{2} \boldsymbol{P},
\end{align}
deduced by taking the $\mathrm{curl}$ of Faraday's law as well as using Gauss' and Ampere's laws in non-magnetic dielectrics, in which there are no free charges and currents. Here $c^{-2} = \mu_{0} \varepsilon_{0}$, and, since matter does not respond instantaneously to stimulation by light, the polarization vector $\boldsymbol{P}$ is related to the electric field $\boldsymbol{E}$ via
\begin{align}
\frac{\boldsymbol{P}}{\varepsilon_{0}} = \chi^{(1)} \circ \boldsymbol{E} + \chi^{(2)} \circ \boldsymbol{E} \otimes \boldsymbol{E} + \chi^{(3)} \circ \boldsymbol{E} \otimes \boldsymbol{E} \otimes \boldsymbol{E} + \ldots, \nonumber
\end{align}
where $\chi^{(i)}$ is the susceptibility tensor of rank $i+1$, symbol $\circ$ stands for a convolution, e.g. $\chi^{(1)} \circ \boldsymbol{E} = \int{\chi^{(1)}(t-\tau_{1}) \boldsymbol{E}(\tau_{1}) \, \d \tau_{1}} \equiv \boldsymbol{P}^{(1)}$, and $\otimes$ for a dyadic product. If the dielectric is centrosymmetric, i.e. $\boldsymbol{E} \rightarrow - \boldsymbol{E}$ implies $\boldsymbol{P} \rightarrow - \boldsymbol{P}$, then $\chi^{(2)}_{jkl} = 0$ and $\chi^{(3)}_{jklm}$ has only three independent components \cite{Newell:1992}. As a result, \eqref{eqn:EM-propagation} becomes
\begin{align}
\label{eqn:Maxwell:wave}
L \boldsymbol{E} = \boldsymbol{P}^{(3)}(\boldsymbol{E}),
\end{align}
where the linear operator is
\begin{align}
L \boldsymbol{E} = \nabla^{2} \boldsymbol{E} - \nabla \left(\nabla \cdot \boldsymbol{E}\right) - \frac{1}{c^{2}} \partial_{t}^{2} \left(1 + \chi^{(1)} \circ\right) \boldsymbol{E}
\end{align}
and the nonlinear term on the rhs is \cite{Newell:1992}
\begin{multline}
\!\!\!\frac{1}{\varepsilon_{0}}\boldsymbol{P}^{(3)} = \int{\chi^{(3)}(t-\tau_{1},t-\tau_{2},t-\tau_{3})} \\ {\left(\boldsymbol{E}(\tau_{1}) \cdot  \boldsymbol{E}(\tau_{2})\right)  \boldsymbol{E}(\tau_{3}) \, \d \tau_{1}\d \tau_{2}\d \tau_{3}}; \nonumber
\end{multline}
here $\chi^{(3)}(t_{1},t_{2},t_{3}) = \chi^{(3)}_{1111}(t_{1},t_{2},t_{3})$.

With the idea to study the evolution of a wavepacket $\boldsymbol{E} = \left[\mathcal{E}(t) e^{\i \left(k z - \omega t\right)} + \mathrm{c.c.}\right] \boldsymbol{e}_{r}$ propagating in the $z$-direction, where $\boldsymbol{e}_{r} \perp \boldsymbol{e}_{z}$ as the light wave is transverse, the linear part of the polarization reduces to
\begin{multline}
\frac{1}{\varepsilon_{0}} \boldsymbol{P}^{(1)}(t) = e^{\i (k z - \omega t)} \Big[\widehat{\chi}^{(1)}(\omega) + \i \dot{\widehat{\chi}}^{(1)}(\omega) \partial_{t} \\
+ \frac{1}{2} \ddot{\widehat{\chi}}^{(1)}(\omega) \partial_{t}^{2} + \ldots\Big] \mathcal{E}(t)  \boldsymbol{e}_{r} + \mathrm{c.c.},
\end{multline}
because the main contribution to the integral summing over frequencies comes from the vicinity of $\omega(k)$. The solution at the leading order can then be represented as
\begin{align}
\boldsymbol{E} \simeq \epsilon \boldsymbol{E}_{0} = \epsilon \left[\mathcal{E}_{0}(T,\tau,R,Z_{1},Z_{2},\ldots) e^{\i \left(k z - \omega t\right)} + \mathrm{c.c.}\right] \boldsymbol{e}_{r}, \nonumber
\end{align}
where the amplitude $\mathcal{E}_{0}$ depends on slow multiple time $T_{i} = \epsilon^{i} t$ and spatial $R = \epsilon r$, $Z_{i} = \epsilon^{i} z$ scales with $i=0,1,2$. Next, we transform the corresponding derivatives to $\partial_{t} \rightarrow \partial_{t} + \epsilon \partial_{T_{1}} + \epsilon^{2} \partial_{T_{2}}, \ \partial_{r} \rightarrow \epsilon \partial_{R}$, $\partial_{z} \rightarrow \partial_{z} + \epsilon \partial_{Z_{1}} + \epsilon^{2} \partial_{Z_{2}}$, expand the linear operator $L = L_{0} + \epsilon L_{1} + \epsilon^{2} L_{2} + \ldots$ and the solution $\boldsymbol{E} = \epsilon \boldsymbol{E}_{0} + \epsilon^{2} \boldsymbol{E}_{1} + \epsilon^{3} \boldsymbol{E}_{2} + \ldots$, so that the wave equation \eqref{eqn:Maxwell:wave} reduces to an iterative system
\begin{subequations}
\begin{align}
\label{eqn:zero-order}
\mathcal{O}(\epsilon)&: & L_{0} \boldsymbol{E}_{0} &= 0, \\
\label{eqn:first-order}
\mathcal{O}(\epsilon^{2})&: & L_{0} \boldsymbol{E}_{1} &= - L_{1} \boldsymbol{E}_{0}, \\
\label{eqn:second-order}
\mathcal{O}(\epsilon^{3})&: & L_{0} \boldsymbol{E}_{2} &= - L_{2} \boldsymbol{E}_{0} - L_{1} \boldsymbol{E}_{1} + \boldsymbol{N}(\boldsymbol{E}_{0}).
\end{align}
\end{subequations}

The leading-order linear operator
\begin{align}
L_{0} = \begin{pmatrix}
      \partial_{z}^{2} - \frac{1}{c^{2}} \partial_{t}^{2} \left[1 + \widehat{\chi}^{(1)}(\omega)\right] & 0 \\
      0 & - \frac{1}{c^{2}} \partial_{t}^{2} \left[1 + \widehat{\chi}^{(1)}(\omega)\right]
    \end{pmatrix}, \nonumber
\end{align}
applied to $\boldsymbol{E}_{0}$ establishes the dispersion relation
\begin{align}
\label{eqn:dispersion-relation}
k^{2}(\omega) = \omega^{2} c^{-2} n^{2}(\omega), \ n^{2}(\omega) = 1 + \widehat{\chi}^{(1)}(\omega).
\end{align}
At the next order, applying
\begin{align}
\label{operator:L-1}
L_{1} = \begin{pmatrix}
          2 \partial_{z} \partial_{Z_{1}} - \frac{1}{c^{2}} \left[\ldots\right] & - \partial_{R} \partial_{z} \\
          - \partial_{R} \partial_{z} - \frac{1}{R} \partial_{z} & - \frac{1}{c^{2}} \left[\ldots\right]
        \end{pmatrix},
\end{align}
to $\boldsymbol{E}_{0}$, where the expression in the square brackets is $\left[\ldots\right] = \i \dot{\widehat{\chi}}^{(1)}(\omega) \partial_{t}^{2} \partial_{T_{1}} + 2 \left[1 + \widehat{\chi}^{(1)}(\omega)\right] \partial_{t} \partial_{T_{1}}^{2}$, we get
\begin{multline}
L_{1} \boldsymbol{E}_{0} = \i k e^{\i (k z - \omega t)} \Big[2 \left(\mathcal{E}_{0 Z_{1}} + \dot{k} \mathcal{E}_{0 T_{1}}\right) \boldsymbol{e}_{r} \\
- \left(\mathcal{E}_{0 R} + \frac{1}{R} \mathcal{E}_{0}\right) \boldsymbol{e}_{z}\Big] + \mathrm{c.c.},
\end{multline}
where we used the fact that $\frac{1}{c^{2}} \left[\omega^{2} \dot{\widehat{\chi}}^{(1)}(\omega) + 2 \omega \left(1 + \widehat{\chi}^{(1)}(\omega)\right)\right] = 2 k \dot{k}$, as follows by differentiating \eqref{eqn:dispersion-relation}. As a result, for the first-order approximation \eqref{eqn:first-order} to have a non-secular solution, we must require that the envelope $\mathcal{E}_{0}$ propagates with the group speed $c_{g} = \frac{\d \omega}{\d k} = \dot{k}^{-1}$ in the $z$-direction:
\begin{align}
\label{condition:no-resonance}
\mathcal{E}_{0 Z_{1}} + \dot{k} \mathcal{E}_{0 T_{1}} = 0,
\end{align}
under which condition the first-order correction becomes
\begin{align}
\label{sln:E-1}
\boldsymbol{E}_{1} = \left[e^{\i (k z - \omega t)} \frac{\i}{k} \left(\mathcal{E}_{0 R} + \frac{1}{R} \mathcal{E}_{0}\right) + \mathrm{c.c.}\right] \boldsymbol{e}_{z},
\end{align}
where we omitted the contribution $\left[e^{\i (k z - \omega t)} \mathcal{E}_{1} + \mathrm{c.c.}\right] \boldsymbol{e}_{r}$ as it can be absorbed in the $\mathcal{E}_{0}$-solution since both correspond to the same null space of $L_{0}$.

The second-order operator takes the form
\begin{align}
\label{operator:L-2}
L_{2} = \begin{pmatrix}
          \partial_{Z_{1}}^{2} + 2 \partial_{z} \partial_{Z_{1}} - \frac{1}{c^{2}} \left[\ldots\right] & - \partial_{R} \partial_{Z_{1}} \\
          - \partial_{R} \partial_{Z_{1}} - \frac{1}{R} \partial_{Z_{1}} & \Delta_{R} - \frac{1}{c^{2}} \left[\ldots\right]
        \end{pmatrix},
\end{align}
where the expression in the square brackets is now
\begin{multline}
\label{eqn:square-brackets:L2}
\!\!\!\!\!\!\!\left[\ldots\right] = \left[\i \dot{\widehat{\chi}}^{(1)}(\omega) \partial_{T_{2}} - \frac{1}{2} \ddot{\widehat{\chi}}^{(1)}(\omega) \partial_{T_{1}}^{2}\right] \partial_{t}^{2} \\
\!\!\!\!\!+ 2 \left[1 + \widehat{\chi}^{(1)}(\omega)\right] \left(2 \partial_{t} \partial_{T_{2}} + \partial_{T_{1}}^{2}\right) + 2 \i \dot{\widehat{\chi}}^{(1)}(\omega) \partial_{t} \partial_{T_{1}}^{2}.
\end{multline}
Applying $L_{2}$ to $\boldsymbol{E}_{0}$ and using \eqref{condition:no-resonance}, we get
\begin{align}
L_{2} \boldsymbol{E}_{0} = e^{\i (k z - \omega t)}
        \begin{pmatrix}
          \left[2 \i k \partial_{Z_{2}} - k \ddot{k} \partial_{T_{1}}^{2}\right] \mathcal{E}_{0} + 2 \i k \dot{k} \mathcal{E}_{0 T_{2}} \\
          - \mathcal{E}_{0 R Z_{1}} - \frac{1}{R} \mathcal{E}_{0 Z_{1}}
        \end{pmatrix}
        + \mathrm{c.c.} \nonumber
\end{align}
after \eqref{eqn:square-brackets:L2} is simplified by differentiating \eqref{eqn:dispersion-relation} twice. Also,
\begin{align}
\!\!\!L_{1} \boldsymbol{E}_{1} = e^{\i (k z - \omega t)}
        \begin{pmatrix}
          \partial_{R} \left[\mathcal{E}_{0 R} + \frac{1}{R} \mathcal{E}_{0}\right] \\
          - 2 k^{2} \dot{k} \partial_{T_{1}} \left[\mathcal{E}_{0 R} + \frac{1}{R} \mathcal{E}_{0}\right]
        \end{pmatrix}
        + \mathrm{c.c.}.
\end{align}
Altogether, the solvability condition removing a resonance from \eqref{eqn:second-order}, i.e. the factor multiplying $e^{\i (k z - \omega t)}$, becomes
\begin{multline}
\label{condition:solvability}
\boldsymbol{e}_{r}: \ \left(2 \i k \partial_{Z_{2}} - k \ddot{k} \partial_{T_{1}}^{2}\right) \mathcal{E}_{0} + 2 \i k \dot{k} \mathcal{E}_{0 T_{2}} \\
+ \partial_{R} \left[\mathcal{E}_{0 R} + \frac{1}{R} \mathcal{E}_{0}\right] + \frac{3 \omega^{2}}{c^{2}} \widehat{\chi}_{3} \mathcal{E}_{0}^{2} \mathcal{E}_{0}^{*} = 0,
\end{multline}
since the term in $\boldsymbol{P}^{(3)}$ proportional to $e^{\i (k z - \omega t)}$ is $- \epsilon^{3} \, 3 \, \omega^{2} c^{-2} \widehat{\chi}_{3} \mathcal{E}_{0}^{2} \mathcal{E}_{0}^{*}$, where $\widehat{\chi}_{3} = \widehat{\chi}_{3}(-\omega,\omega,\omega)$ \cite{Ablowitz:2011}.

To bring \eqref{condition:solvability} to a canonical form, we first cross-differentiate \eqref{condition:no-resonance} to show that $\mathcal{E}_{0 Z_{1}Z_{1}} = \dot{k}^{2} \mathcal{E}_{0 T_{1}T_{1}}$, and, second, transform to the frame of reference moving with the group velocity $\left(T_{2}, Z_{2}\right) \rightarrow \left(T_{2}^{\prime} = T_{2}, Z_{2}^{\prime} = Z_{2} - \dot{k}^{-1} T_{2}\right)$, thus eliminating the along-the-beam coordinate $Z_{2}$. Note that the slow time scale $T_{2}$ naturally appears at $\mathcal{O}(\epsilon^{3})$ in \eqref{eqn:second-order} and is independent of $T_{1}$. The resulting equation can be put in a non-dimensional form
\begin{align}
\label{NLS:canonical:non-dimensional}
\i \psi_{\tau} - \alpha \psi_{zz} + \left(\Delta_{r} - r^{-2}\right) \psi + \beta |\psi|^{2} \psi = 0,
\end{align}
where $\alpha = \sgn{\ddot{k}}$, $\beta = \widehat{\chi}_{3}/|\widehat{\chi}_{3}|$, and we scaled the independent and dependent variables:
\begin{align}
R \rightarrow \frac{r}{k}, \ Z_{1}^{\prime} \rightarrow \left|\frac{\ddot{k}}{k \dot{k}^{2}}\right|^{1/2} \! z, \ T_{2}^{\prime} \rightarrow \frac{2 \dot{k}}{k} \tau, \
\mathcal{E}_{0} \rightarrow \frac{k c}{\omega} \frac{\psi}{\sqrt{3 |\widehat{\chi}_{3}|}}. \nonumber
\end{align}
In most of the earlier cited works the dependence on $Z_{1}$, i.e. the term $- \alpha \psi_{zz}$ in \eqref{NLS:canonical:non-dimensional}, is omitted, which corresponds to the situation when the duration and transverse width of the light pulses are such that the modulation can be treated as stationary \cite{Luther:1994}, thus simplifying \eqref{NLS:canonical:non-dimensional} to
\begin{align}
\label{GP:general}
\i \psi_{\tau} + \left(\Delta_{r} - r^{-2}\right) \psi + \beta |\psi|^{2} \psi = 0,
\end{align}
where $\beta=\pm 1$ if we neglect absorption in susceptibility. Equation \eqref{GP:general} is of Gross-Pitaevskii (GP) type, originally derived \cite{Gross:1961,*Pitaevskii:1961} to describe the ground state of a Bose-Einstein condensate in an external potential $V$. While the appearance of a potential $\sim r^{-2}$ is generic for cylindrical envelope wave equations, the factor $-1$ vs $-\frac{1}{4}$ in the heuristic analysis leading to \eqref{eqn:SE:cylindrical} comes from the geometric setting: \eqref{condition:solvability} is derived in a paraxial approximation in 3D \cite{Leontovich:1944}, while the heuristic analysis was done in 2D. Same as NLS, \eqref{GP:general} is derived for small, but finite, amplitude $\epsilon$ solutions based on a balance of nonlinearity and dispersion of the wavepacket, which occurs only at some distance $r$ from the origin as the wave amplitude varies with $r$ -- this is a crucial difference from the translationally invariant case of NLS on a line, in which one can take the limit of small amplitude solutions and be left with the same linear part; in the cylindrical case this is no longer the case, i.e. the linear part of \eqref{GP:general}, in which nonlinearity and dispersion are balanced, does not correspond to \eqref{eqn:SE:cylindrical}, when nonlinearity is absent.

Notably, an inverse square potential also arises in quantum mechanics \cite{Case:1950,*Kalf:1975,*Reed:1979}, molecular physics \cite{Camblong:2001}, nuclear physics \cite{Beane:2001,*Esteve:2002}, black holes \cite{Regge:1957,*Zerilli:1970,*Moncrief:1974,*Strominger:1998,*Claus:1998,*Azcarraga:1999,*Solodukhin:1999,*Michelson:2000,*Papadopoulos:2000,*Bellucci:2002,*Carlip:2002},
in wave propagation on conic manifolds \cite{Cheeger:1982}, and in the theory of combustion \cite{Bebernes:1989}. Since the Laplacian and the inverse square potential are of equal strength \footnote{Because of that it is known to have some peculiar properties such as no ground state, i.e. there is no lower limit on the allowed energies \cite{Essin:2006}, symmetry breaking anomaly in the process of renormalization \cite{Essin:2006,*Camblong:2000,*Coon:2002}.}, the latter cannot be neglected. For the same reason, the GP equation \eqref{GP:general} retains the scaling symmetry $\psi(\tau,r) \mapsto a \, \psi(a^{2} \tau, a \, r)$, which can be used to generate solutions. Equation \eqref{GP:general} admits a mechanistic interpretation by looking for a solution in the form $\psi = a \, e^{\i S(r)}$, thus leading to the Hamiltonian $p^{2} + r^{-2} = a^{2} \equiv H$, where $p = S_{r}$ is the momentum and we omitted the terms $S_{rr} + \frac{1}{r} S_{r}$ as in the WKB method, which is to be verified a posteriori. Clearly, the potential is repulsive and hence trajectories must tend to infinity; therefore, $S_{r} \approx a - \frac{1}{2 a r^{2}}$ for large $r$ and neglecting the omitted terms is justified. Introducing an ``effective'' time $t$, the corresponding Hamiltonian system reduces to $\ddot{r} = \frac{4}{r^{3}}$, integration of which indeed gives diverging trajectories corresponding to the solutions of \eqref{GP:general} oscillatory at infinity.

\begin{figure*}
	\setlength{\labelsep}{-5.0mm}
	\centering
    \sidesubfloat[]{\includegraphics[height=1.325in]{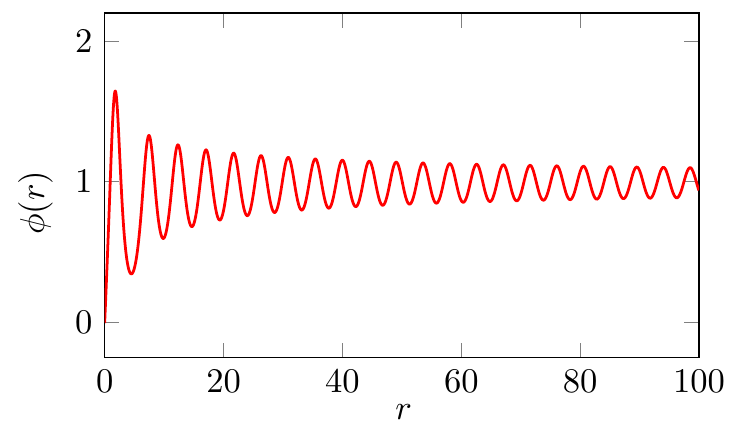}\label{fig:plot-u-regularplus}}
	\sidesubfloat[]{\includegraphics[height=1.325in]{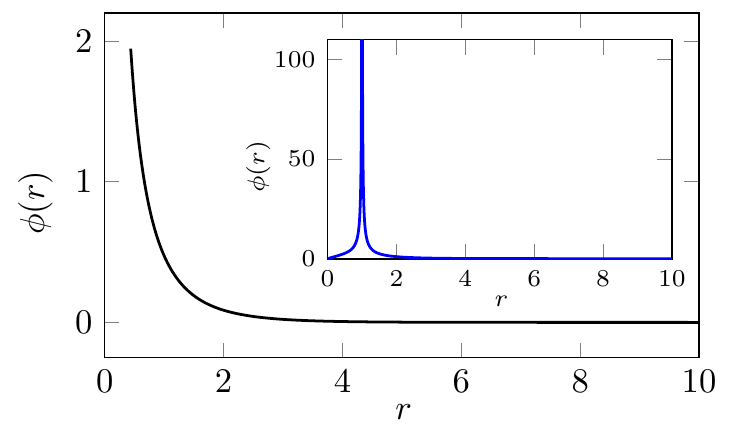}\label{fig:plot-u-singularplus}}
	\sidesubfloat[]{\includegraphics[height=1.325in]{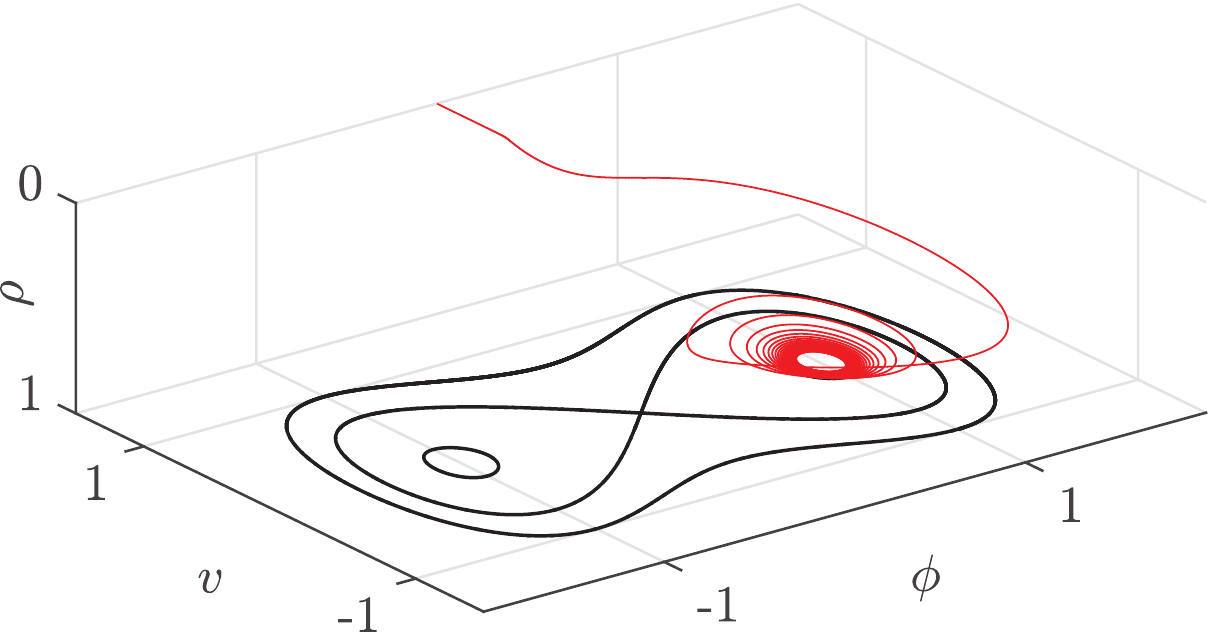}\label{fig:trajectory}}
\caption{Real solutions to \eqref{eqn:Stokes-wave:canonical}: (a) regular, $\beta = 1$, (b) singular at the origin, $\beta = 1$; inset shows a solution singular on the ring $r_{0}=1$, $\beta = - 1$, (c) trajectory of \eqref{system:Stokes-wave} corresponding to solution in Fig.~\ref{fig:plot-u-regularplus}; phase portrait of \eqref{system:Stokes-wave} at $\rho = 1$.} \label{fig:Stokes-solutions}
\end{figure*}
While it is not the purpose to explore all possible solutions of \eqref{GP:general}, let us look for standing wave ground states $\psi(\tau,r) = e^{\i \mu \tau} \phi(r)$ known as ``breathers''; when $\mu > 0$ without loss of generality we can take $\mu=1$ since it can be scaled out of \eqref{GP:general} producing for complex $\phi$
\begin{align}
\label{eqn:Stokes-wave:canonical}
- \phi + \left(\Delta_{r} - r^{-2}\right) \phi + \beta \, |\phi|^{2} \phi = 0.
\end{align}
While \eqref{eqn:Stokes-wave:canonical} is not integrable analytically, we can determine asymptotics of its solutions:
\begin{subequations}
\begin{align}
\label{asymptotics:origin}
r \rightarrow 0&: \ \phi \sim r, \ \frac{A}{r \, (\ln{r})^{1/2}} \ \text{with} \ A^{2} = - \frac{1}{\beta}; \\
\label{asymptotics:infinity}
r \rightarrow \infty&: \ \phi \sim r^{-1/2} e^{-r + \frac{3}{8 r}}, \ 1 + \const \, e^{- 2 \sqrt{-\beta} r}.
\end{align}
\end{subequations}
As opposed to the radial NLS \eqref{NLS:basic:non-dimensional} -- the asymptotics of solutions of which near the origin includes only $\phi \sim \const$ instead of \eqref{asymptotics:origin} -- in addition to the solutions regular at the origin $\sim r$ \eqref{asymptotics:origin} corresponding to the focusing case $\beta = 1$, cf. Fig.~\ref{fig:plot-u-regularplus}, equation \eqref{eqn:Stokes-wave:canonical} admits $\phi(r)$ singular at the origin \eqref{asymptotics:origin} \cite{Note1} as illustrated in Fig.~\ref{fig:plot-u-singularplus}, also for $\beta = 1$. In the defocusing case $\beta = -1$, \eqref{eqn:Stokes-wave:canonical} has solutions with asymptotics $\phi \sim C \, r^{-1/2} (r-r_{0})^{-1}$ singular along a ring of radius $r_{0} = \frac{1}{2} C^{2}$, cf. inset in Fig.~\ref{fig:plot-u-singularplus}. The solitons in Fig.~\ref{fig:plot-u-singularplus} are ``bright'', i.e. localized in space and evanescent at infinity, regardless of the sign of $\beta$, while in Fig.~\ref{fig:plot-u-regularplus} the soliton is ``dark''. Despite the singular nature of the solitons in Fig.~\ref{fig:plot-u-singularplus}, they are as valuable as the widely studied finite-time singularity of \eqref{NLS:basic:non-dimensional}, being indicative of a localized behavior in the unreduced physical system such as the Maxwell equations.

\footnotetext[1]{this solution is analogous to the lowest-order solution (without nodes in the radial profile) of the radial NLS found numerically by Chiao et al. \cite{Chiao:1964}}

Recasting \eqref{eqn:Stokes-wave:canonical} as an autonomous system of first-order equations, in which a singularity at the origin is removed by introducing new independent $\xi = r - \frac{1}{r} + 2 \ln{r} \in (-\infty,+\infty)$ and dependent $\rho = r / (r+1) \in [0,1]$ variables, we may understand the solution variety of \eqref{eqn:Stokes-wave:canonical} through a dynamical systems point of view \cite{Jones:1986,*Newton:1993}:
\begin{subequations}
\label{system:Stokes-wave}
\begin{align}
\dot{\phi} &= \rho^{2} v, \\
\dot{v} &= - \rho \, (1-\rho) v + (1-\rho)^{2} \phi + \rho^{2} (\phi - \beta \phi^{3}), \\
\dot{\rho} &= \rho^{2} (1-\rho)^{2},
\end{align}
\end{subequations}
where $\dot{( \ )} = ( \ )_{\xi}$. We find that all solutions starting at $\rho=0$ end up being attracted to one of the trajectories in the invariant plane $\rho=1$ such as in Fig.~\ref{fig:trajectory} corresponding to the solution in Fig.~\ref{fig:plot-u-regularplus}; the phase portrait at $\rho=1$ exhibits a saddle point at the origin, two homoclinic orbits passing through it, and two centers at $(\phi,v)=(\pm1,0)$, thus separating domains of solution's attraction.

Same as NLS, \eqref{GP:general} is of Hamiltonian type. To analyze its conservation laws, let us supply the initial-value problem for \eqref{GP:general} with the boundary conditions:
\begin{align}
\label{BCs:cNLS}
r=0: \ \psi_{r} = 0; \ r \rightarrow \infty: \ \psi \rightarrow 0.
\end{align}
We also naturally assume that at $r=0$ the solution itself is non-singular as well as decays sufficiently fast as $r \rightarrow \infty$ \footnote{This should be valid at least initially if the initial condition is chosen as a compact/localized perturbation of finite energy}, so that the deduced below integral conservation laws are well-defined. Multiplying \eqref{GP:general} with $\overline{\psi} = \psi^{r} - \i \psi^{i}$, taking the imaginary part, and integrating over the cylindrical measure $\int{\d \nu} = 2 \pi \int_{0}^{\infty}{r \, \d r}$ leads to the conservation of the number of particles $\mathcal{N} = \int{|\psi|^{2} \, \d \nu}$ defined so in analogy to quantum mechanics, which is the consequence of the invariance of \eqref{GP:general} under the phase-shift. Similarly, multiplying \eqref{GP:general} with $\overline{\psi}_{\tau}$, taking the real part of the resulting expression, and integrating over the cylindrical measure we get the conservation of the Hamiltonian $\mathcal{H} = \frac{1}{2} \int{\left[|\psi_{r}|^{2} + \frac{1}{r^{2}}|\psi|^{2} + \frac{\beta}{2} |\psi|^{4}\right] \, \d \nu}$. Hence, equation \eqref{GP:general} admits a canonical Hamiltonian formulation:
\begin{align}
J U_{t} = \frac{\delta \mathcal{H}}{\delta U}, \ \text{where} \ J = \begin{pmatrix}
                                                            0 & -1 \\
                                                            1 & 0
                                                          \end{pmatrix}, \
                                                      U = \begin{pmatrix}
                                                            u \\
                                                            v
                                                          \end{pmatrix}.
\end{align}
and $\psi = u + \i v$. Assuming that after integration by parts all boundary terms do not contribute (e.g. at $r=0$ due to the solution being symmetric $U_{r}=0$ or due to vanishing variation $\delta U$ thus fixing the space of allowed perturbations), we find for the first variation:
\begin{align}
\delta \mathcal{H} = \int{\left[\frac{U}{r^{2}} - \frac{1}{r}\frac{\partial}{\partial r}\left(r\frac{\partial U}{\partial r}\right) - \beta \, |U|^{2} U\right] \cdot \delta U \d\nu},
\end{align}
where dot denotes scalar product. Obviously, the standing wave solution $\phi(r)$ is a fixed point of a constrained Hamiltonian $\mathcal{H} + \lambda \mathcal{N}$, i.e. \eqref{eqn:Stokes-wave:canonical} is $\frac{\delta \mathcal{H}}{\delta U} + \lambda \frac{\delta \mathcal{N}}{\delta U} = 0$ with the Lagrange multiplier $\lambda=\frac{1}{2}$. Formal stability of $\phi(r)$ is gleaned from the second variation $\delta^{2} \mathcal{H} + \lambda \, \delta^{2} \mathcal{N}$ of the constrained Hamiltonian, where $\delta^{2} \mathcal{N} = \int{\delta U \cdot \delta U \, \d \nu}$ and
\begin{align}
\begin{split}
\delta^{2} \mathcal{H} = \frac{1}{2} \int \Bigl\{\delta U_{r} \cdot \delta U_{r} + \frac{\delta U \cdot \delta U}{r^{2}} - \beta \big[4 u v \, \delta u \delta v\\
+\left(3 u^{2} + v^{2}\right) (\delta u)^{2} + \left(3 v^{2} + u^{2}\right) (\delta v)^{2}\big]\Bigl\} \, \d\nu;
\end{split}
\end{align}
however, only allowed variations, namely tangent to the constraint $\delta \mathcal{N} = 0$, that is $U \cdot \delta U = 0$, can be considered \cite{Maddocks:1993}, thus reducing the Hessian density of $\delta^{2} \mathcal{H} + \lambda \, \delta^{2} \mathcal{N}$ to
\begin{align}
(\delta u)^{2}: \ \left(\frac{1}{2 r^{2}} + \lambda\right) \left(1+\frac{u^{2}}{v^{2}}\right) - \frac{\beta}{2} \left[2 u^{2} + v^{2} + \frac{u^{4}}{v^{2}}\right]. \nonumber
\end{align}
Therefore, the potential in the GP equation \eqref{GP:general} plays a stabilizing role; in the limit of large $r$ the definiteness of the Hessian is determined by the sign of $\beta$ -- with $\beta=1$ implying instability such as for the solution in Fig.~\ref{fig:plot-u-regularplus} and $\beta=-1$ stability \footnote{The stability implications in the defocusing case are, of course, formal in the sense that positive-definiteness of the constrained Hamiltonian is not a sufficient condition for a local minimum to occur in infinite dimensions.} -- thus recovering the known fact that solutions of the focusing 1D NLS are unstable \cite{Zakharov:1968}, in particular, leading to a finite-time singularity when nonlinearity overpowers the dispersive spreading.

To identify the conditions for the latter phenomena, one can derive an equation for the evolution of a variance $\mathcal{V}(\tau) = \int{r^{2} |\psi|^{2} \, \d \nu}$ in analogy to that for NLS  \cite{Vlasov:1971,Fibich:2015}:
\begin{align}
\label{variance:derivative:second:final}
\frac{1}{4} \frac{\d^{2} \mathcal{V}}{\d \tau^{2}} = 4 \, \mathcal{H} + 2 \pi \gamma |\psi(\tau,0)|^{2},
\end{align}
where $\gamma$ labels the coefficient originating from the potential term $V(r) = \frac{\gamma}{r^{2}}$ in \eqref{GP:general}. Integrating \eqref{variance:derivative:second:final} yields
\begin{multline}
\mathcal{V}(\tau) = \mathcal{V}(0) + \mathcal{V}^{\prime}(0) \tau + 8 \mathcal{H} \tau^{2} \\
+ 8 \pi \gamma \int_{0}^{\tau}{\d \tau^{\prime}\int_{0}^{\tau^{\prime}}{|\psi(\tau^{\prime\prime},0)|^{2} \d \tau^{\prime\prime}}},
\end{multline}
Should $\gamma=0$ as in the case of the radial NLS \eqref{NLS:basic:non-dimensional}, and if the initial conditions are such that $\mathcal{H} < 0$, i.e. $\mathcal{V}^{\prime\prime}(0) = 16 \mathcal{H} < 0$, then from the solution of the quadratic equation $\frac{1}{2} \mathcal{V}^{\prime\prime}(0) \tau_{*}^{2} + \mathcal{V}^{\prime}(0) \tau_{*} + \mathcal{V}(0) = 0$ it follows
\begin{align}
\tau_{*} = \left(- \mathcal{V}^{\prime}(0) + \sqrt{\mathcal{V}^{\prime 2}(0) - 2 \mathcal{V}(0) \mathcal{V}^{\prime\prime}(0)}\right)/\mathcal{V}^{\prime\prime}(0),
\end{align}
where necessarily $\mathcal{V}(0)>0$ and $\mathcal{V}^{\prime}(0)<0$; thus there exists a finite time $\tau_{*} > 0$ such that $\mathcal{V} \rightarrow 0$ in contradiction to its definition, which shows that it has to be positive. The $H^{1}$-solution must therefore develop a singularity $|\psi| \rightarrow \infty$, $|\psi_{r}| \rightarrow \infty$ at $r \rightarrow 0$ no later than the time $\tau_{*}$ \cite{Glassey:1977,Note3}. However, the presence of the potential leads to an extra term $2 \pi \gamma |\psi(\tau,0)|^{2}$ in \eqref{variance:derivative:second:final}: if $\gamma < 0$ as in our case then, since the integral of $|\psi(\tau,0)|^{2}$ is positive-definite, the finite-time singularity still takes the place, while for $\gamma > 0$ the situation may potentially change and prevent the singularity from formation altogether, i.e. if the growth of the last term in \eqref{variance:derivative:second:final} with time is faster than $- 8 \mathcal{H} \tau^{2}$.

\footnotetext[3]{This means that the solution gets out of the $H^{1}$-space, so that the condition that $\mathcal{V}$ must be positive (when $\psi \in H^{1}$) does not need to be satisfied any longer.}

%


\end{document}